\documentstyle[12pt,moriond,epsf]{article}

\newcommand{\arcsec}{$''$}
\newcommand{\um}{$\mu$m}
\newcommand{\funit}{$\, \rm mJy \, arsec^{-2}$}
\newcommand{\halpha}{H$\alpha$}
\begin{document}
%
\noindent
To appear in the Proceedings of the XVII$^{\rm th}$ Moriond
Astrophysics Meeting:\\``Extragalactic Astronomy in the Infrared''
\medskip
\hrule
\vskip -5mm
\heading{First Results from ISOCAM Images of Circumnuclear Starbursts in
Barred Galaxies
\footnote{ This work is based on observations with ISO, an ESA project
with instruments funded by ESA Member States (especially the PI
countries: France, Germany, the Netherlands and the United Kingdom)
and with the participation of ISAS and NASA.}}

\author{H. Wozniak $^{1}$, D. Friedli $^{2,3}$, L. Martinet $^{3}$, D.
Pfenniger $^{3}$}
{$^{1}$ Observatoire de Marseille, F-13248 Marseille, France.}
{$^{2}$ D\'epartement de Physique and OMM, Universit\'e Laval,
Ste-Foy, QC, G1K~7P4, Canada.}
{$^{3}$ Observatoire de Gen\`eve, CH-1290 Sauverny, Switzerland.}

\begin{moriondabstract}
\noindent
For a sample of ten nearby barred galaxies, images at 6.75\um\ and
15\um\ of circumnuclear starbursts are being taken by ISOCAM on board
of the ISO satellite.  We first recall the main goals of our project,
give its current status, and finally, as an example, preliminary
results for NGC~4321 (M100) are presented.
\end{moriondabstract}

\section{Introduction}
Gravitational perturbations of galaxies are likely to play an
important role in generating galaxy starbursts and in fueling active
galactic nuclei (AGN). Whereas tidal forces between interacting or
merging galaxies are strong enough to drive major gas flows, the
compression of which leads to intense starbursts and/or activity in
the nuclei, it is less clear whether forming bars alone are also able
to trigger enhanced star formation either along the bar or near the
centre \cite{SAAS97}.  NGC 7479 is an example where this activity can
be significantly high, the strength of the bar seeming to be an
essential parameter to consider \cite{MF97}.  Therefore until recently
the link between bar and starburst has not been very clear (see
\cite{HHG96} and references therein).

The difficulty with ordering these processes is that they occur on
widely different times scales, among which we mention: the bar growth
time, the bar life time at its maximum strength, the time scale for
driving the gas to the centre, the molecular cloud formation time, the
gas cooling time, and finally all the time scales associated with
stellar formation and evolution, such as the energy input from massive
stars, the mass return to the interstellar medium from red giants, and
the metal enrichment.  The involved physics is so complex and some
parts so poorly known that only a comparison between simulations and
high resolution and multi-spectral observations of the central regions
of galaxies can bring some enlightenment on these questions.

One severe issue for feeding AGN is how to accrete gas down to the
centre at subparsec scale, which requires a process able to evacuate
nearly all the angular momentum.  It seems that the same trick to
extract angular momentum working well at kpc scale, the bar, can be
reused at smaller scales, leading to the often observed secondary bars
embedded in the larger one \cite{FM93}.  Such secondary bars can be
observed in the visible \cite{BC93,W95}, NIR \cite{S93,F96,C97} and CO
(\cite{K96} and references therein).

Recent star formation needs also a quantitative description. Several
attempts have been made to relate the star formation rates to
luminosities like e.g. $L_{\rm H\alpha}$, $L_{\rm B}$ and $L_{\rm
FIR}$.  However, $L_{\rm H\alpha}$, related to recent star formation
($10^6$ to $10^7$~yr), is very sensitive to dust absorption while
$L_{\rm B}$ is more related to $1-3 \, \rm M_\odot$ main sequence
stars and thus traces star formation over longer time scales ($4 \cdot
10^8$ to $6 \cdot 10^9$~yr). Dust might not only be heated by young
massive stars but also by older stars so that $L_{\rm FIR}$ (estimated
from 60\um\ and 100\um\ IRAS measurements) remains a controversial
tracer of recent star formation \cite{ST92}.

MIR wavelengths (5--17\um) trace the hot dust associated with the most
recent star formation \cite{TDW93}, and thus might give powerful
indicators of star formation activity. Indeed, broad band imaging
allows to distinguish between the various sources of MIR luminosity
(PAHs, graphite grains, etc.).  Location of sources of heating can be
done at shorter wavelengths.  Moreover, MIR imaging operate with
larger arrays and better resolution than those available so far in the
FIR allowing thus to investigate in a much more detailed way the
intricate connections between global dynamics and star formation
activity.

\section{Project aims}
The broad aim of our ISO program is to clarify the intimate links
between bar dynamics and star formation activity in barred
galaxies. MIR imaging is indeed particularly powerful for locating the
youngest star forming regions and for exploring broader issues like
the dynamical origin of these activities.  Among the sites of high
star formation activity, the nuclear and/or circumnuclear regions have
particularly hold our attention.  We thus decided to observe the
emission from PAHs and hot dust with the highest pixel field of view
(PFOV$=$1.5\arcsec px$^{-1}$) allowed by the ISOCAM camera of the ISO
satellite. The small field of view (45\arcsec) is sufficient to
enclose the circumnuclear star forming regions of nearby barred
galaxies.

Specific goals include: the analysis of PAHs and hot dust
distributions and patterns, the study of gradients in the PAHs/dust
ratio, the emphasis of spatial correlations or offsets between various
physical components (e.g. young and old stars, molecular and ionized
gas, PAHs, dust), the comparison of star formation rates estimated
from MIR emission with other indicators, etc.

\section{Sample and observations}
We have selected a sample of ten nuclear and circumnuclear starburst
galaxies for which low resolution MIR (e.g. \cite{TDW93}) and/or
millimeter maps have been done (cf. Table~1).  All these galaxies show
signatures generally associated with prominent star formation activity,
like \halpha\ emission (e.g. HII regions, hot spots), dust lanes,
mini-spirals, nuclear bar and/or ring, etc.

The ongoing observations (Table~1) were done with the LW2 (6.75\um)
and LW3 (15\um) broad band filters at the 1.5\arcsec px$^{-1}$ PFOV.
The beamswitching mode with 2 reference fields was used to remove the
MIR background. This program requires roughly 12 hours. Its main
difficulty is to deconvolve the image PSF in order to get the best
details which should then be compared with the millimetric
interferometric maps and data at various other wavelengths. Further
details on the reduction and calibration processes can be found in
\cite{WFMP97}.

The LW2 filter (5--8.5\um) includes emission from the PAH bands at
6.2\um, 7.7\um\ and 8.6\um\ as well as the underlying continuum. The
LW3 filter (12--18\um) collects continuum emission of small grains as
well as [NeII] (12.8\um) and [NeIII] (15.5\um) emissions if present.

\begin{table}
\caption{Galaxy sample}
\begin{center}
\begin{tabular}{lll}
Objects & Morphological type & Characteristics \cr
\noalign{\medskip}
{\em Observed (until 31 May 1997)\/}:& & \cr
\hline\noalign{\smallskip}
NGC~1097 & SB(s)b       & Seyfert; \halpha\ ring \cr
NGC~4321 & SAB(s)bc     & Mild starburst; \halpha\ ring \cr
NGC~5236 & SAB(s)c      & Starburst; \halpha\ ring \cr
NGC~7469 & (R')SAB(rs)a & Seyfert; Starburst \cr
NGC~7479 & SB(s)c       & LINER; Starburst \cr
NGC~7552 & (R')SB(s)ab  & Starburst \cr
\noalign{\medskip}
{\em Pending\/}: & & \cr
\hline\noalign{\smallskip}
NGC~1808 & (R)SAB(s)a   & Seyfert \cr
NGC~2903 & SAB(rs)bc    & Starburst \cr
NGC~3351 & SB(r)b       & Starburst; \halpha\ ring \cr
NGC~4691 & (R)SB(s)0/a  & \halpha\ ring \cr
\end{tabular}
\end{center}
\end{table}

\section{A puzzling example: NGC~4321 (M100)}
NGC~4321 (M100) is a Virgo giant moderately barred spiral galaxy. Its
inner region has been extensively studied at almost all wavelengths
(see \cite{K95,S95} and references therein).  Optical, \halpha, NIR
and CO images have revealed many intricate features in the central kpc
associated to its mild star formation activity.  Several HII regions
lie along a 1~kpc circumnuclear ring which crosses a bar 8~kpc long.
A nuclear bar, nearly parallel to the large-scale bar, occupies the
region inside the ring. In fact, this galaxy certainly represents one
of the best laboratory for exploring the interplay between starburst
activity and non-axisymmetric dynamics.

At 6.75\um\ and 15\um, the emission is clearly dominated by two bright
spots located near the nuclear bar ends (Fig.~1). These peaks are
almost twice brighter than the remainder of the circumnuclear
region. A third local maximum, much less intense, is visible at the
nucleus.  Without any deconvolution, firm morphological differences
between LW2 and LW3 images cannot be displayed. Furthermore, the
bar-like feature which seems to connect the two peaks and the nucleus
must be confirmed on deconvolved images.  \cite{WFMP97} have performed
a detailed analysis of these ISOCAM images and have compared them with
U, B, R, \halpha\ and CO data.  Their main findings are summarized
below.  A very schematic view of the central multi-wavelength
morphology is given in Fig.~2.

\vfill

\begin{figure}
\vskip -10truemm
\centerline{\vbox{\epsfysize=120truemm \epsfbox{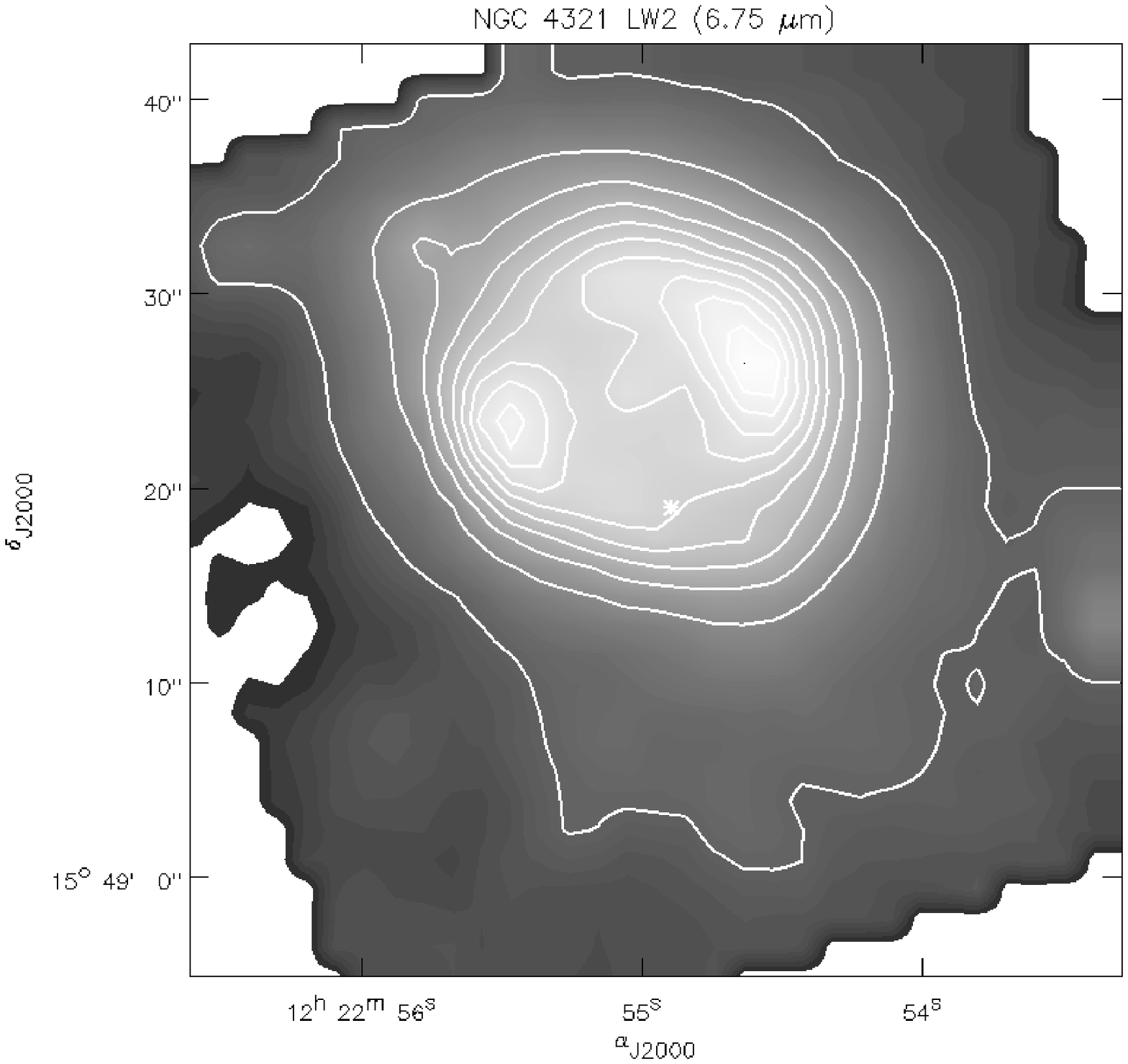}}}
\vskip -10truemm
\centerline{\vbox{\epsfysize=120truemm \epsfbox{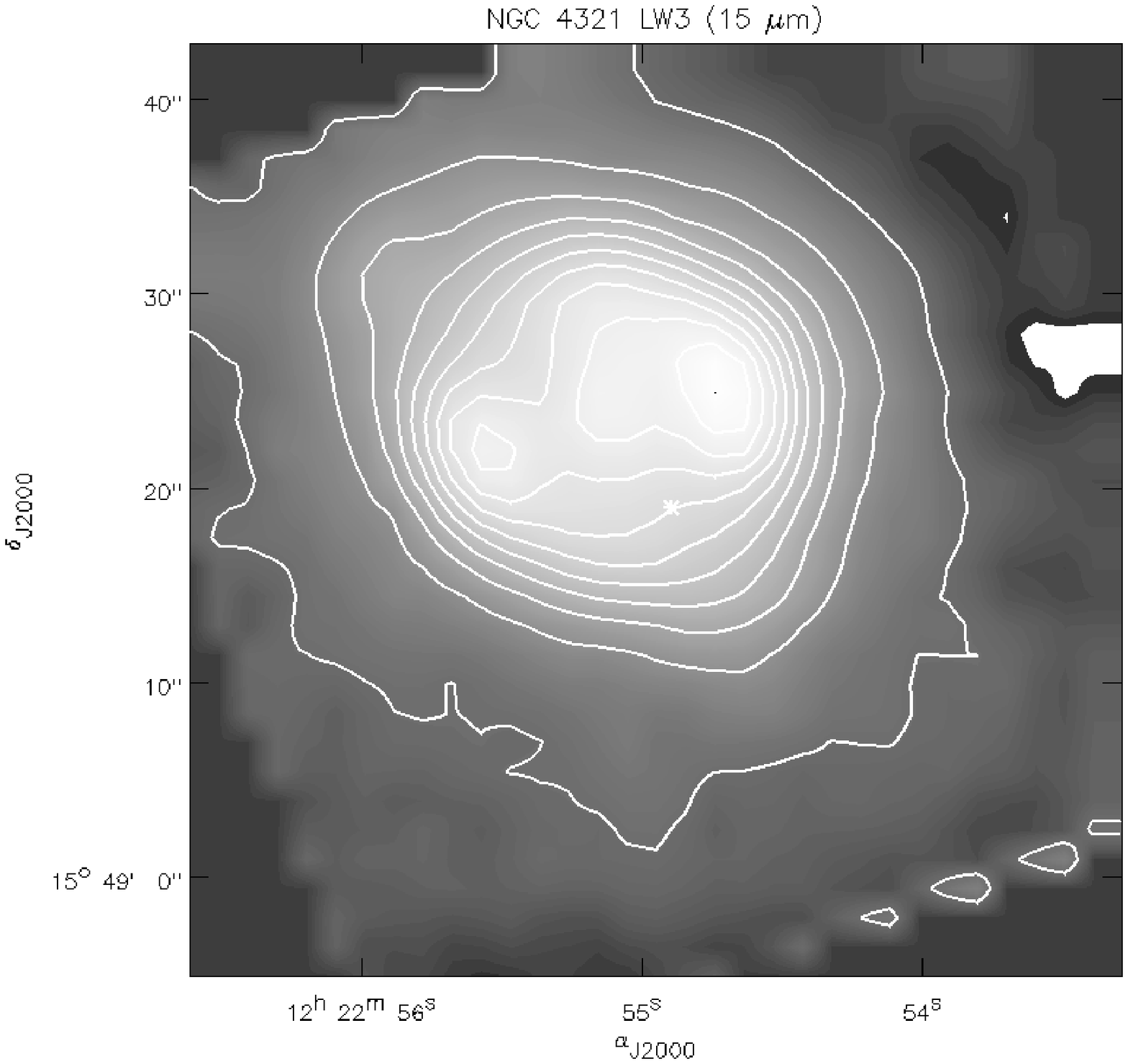}}}
\vskip -5truemm
\caption{Linear greyscale images of NGC~4321 (M100) at 6.75\um\ ({\em
top panel\/}) and 15\um\ ({\em bottom panel\/}). Isocontours spaced by
0.13\funit\ at 6.75\um\ and 0.12\funit\ at 15\um\ are overplotted. The
faintest levels are 0 in both images.  The highest ones are
1.33\funit\ at 6.75\um\ and 1.20\funit\ at 15\um. The PFOV is
1.5\arcsec px$^{-1}$. The effective field of view is 45\arcsec. The
images are not deconvolved.}
\end{figure}

1) The two bright MIR (i.e. LW2 and LW3) peaks at the nuclear bar ends
approximately coincide with strong \halpha\ and K-band emissions, dust
lanes obscuring optical emission (from U- to I-bands), and similar
peaks in CO emission.  Moreover, LW2/LW3$>$1 in these regions which
may mean that PAHs dominate the MIR luminosity.

2) With respect to the nuclear bar (counter-clockwise rotation), the
two MIR maxima are trailing whereas the two \halpha\ maxima are
leading.

3) The two other regions of star formation (visible in U, B and
\halpha\ along the nuclear bar minor axis) are not embedded in dust
lanes, are offset from the CO spiral arms, and seems not to be
associated (or only marginally), neither with a strong emission in
NIR, nor with LW2 and LW3 bands. In these regions, the hot dust
dominates the MIR emission as LW2/LW3$<$1.

4) There are no ``holes'' in the LW2 and LW3 surface brightness,
i.e. emission from PAHs and hot dust is observed everywhere in the
central region.  Of course, this might be due to the map resolution
which induces a 200~pc wide smoothing.

\begin{figure}
\vskip -10truemm
\centerline{\vbox{\epsfysize=130truemm \epsfbox{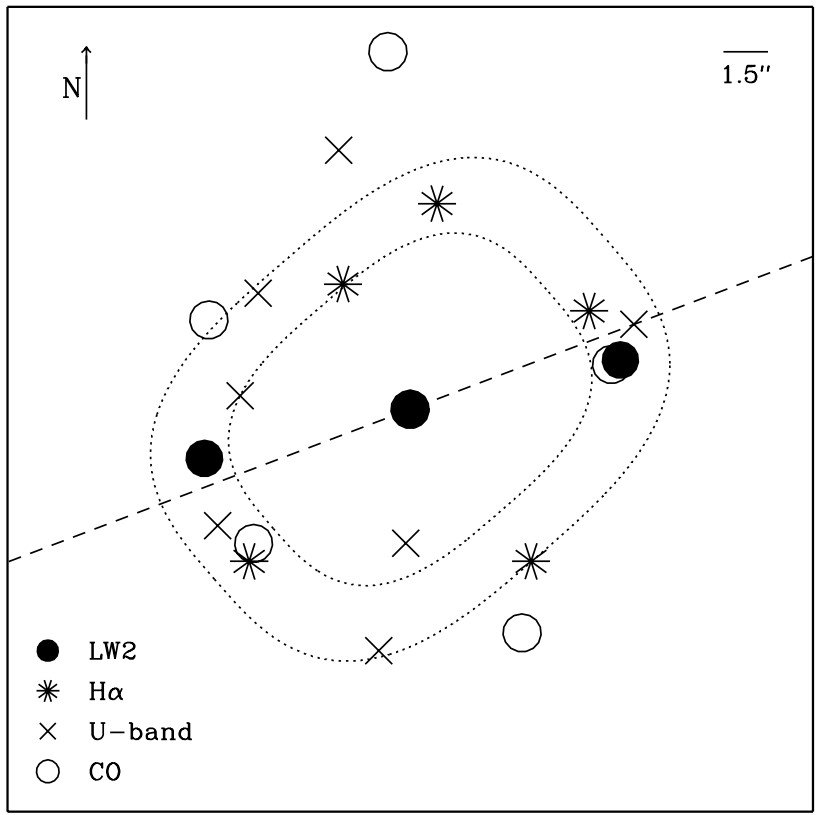}}}
\vskip -10truemm
\caption{Very schematic sketch of the intricate central morphology of
NGC~4321 (M100) including the three LW2 peaks, the direction of the
nuclear bar (dashed line; PA$=111^\circ$ [9]), the position of i) the
nuclear \halpha\ ring (dotted lines; PA$=131^\circ$) with its five
most prominent peaks, ii) five regions with intense CO emission, and
iii) seven bright regions in U-band.}
\end{figure}

\acknowledgements{We are grateful to Dr. A. Zavagno for enlightening
discussions on PAHs.  The project is supported by the Swiss National
Science Foundation (FNRS) through an ``Advanced Researcher''
fellowship to D.F. and the grant 20-49239.96 to L.M. and D.P.
Figure~2 makes use of data kindly provided by F.~Bresolin (U-band),
P.~Martin and J-R.~Roy (\halpha), and R.~Rand (CO).}


\end{document}